\documentclass[pra,twocolumn,showpacs]{revtex4}
\usepackage{epsfig}
\usepackage{amsbsy}
\usepackage{amsmath}
\usepackage{graphicx}
\usepackage{latexsym}

\begin{document}

\title{Disentanglement in a quantum critical environment%
}
\author{Zhe Sun}
\author{Xiaoguang Wang}
\email{xgwang@zimp.zju.edu.cn}
\affiliation{Zhejiang Institute of
Modern Physics, Department of Physics, Zhejiang University, Hangzhou
310027, China}
\author{C. P. Sun}
\email{suncp@itp.ac.cn}
\affiliation{Institute of Theoretical Physics, Chinese Academy of Sciences, Beijing,
100080, China}
\date{\today }

\begin{abstract}
We study the dynamical process of disentanglement of two qubits and
two qutrits coupled to an Ising spin chain in a transverse field,
which exhibits a quantum phase transition. We use the concurrence
and negativity to quantify entanglement of two qubits and two
qutrits, respectively. Explicit connections between the concurrence
(negativity) and the decoherence factors are given for two initial
states, the pure maximally entangled state and the mixed Werner
state. We find that the concurrence and negativity decay
exponentially with fourth power of time in the vicinity of critical
point of the environmental system.
\end{abstract}

\pacs{05.40.-a, 03.65.Yz, 32.80.-t,03.67.Mn} \maketitle

\section{Introduction}

Entanglement is one of the most essential features in quantum mechanics~\cite%
{Ein} and in recent decades has been focused by people in many
fields of physics. Motivated by the progress of quantum information,
entanglement has become a basic resource in the quantum technologies
such as quantum teleportation and quantum
cryptography~\cite{Nielsen}-\cite{Ekert}. On the other hand,
generally a realistic system is surrounded by an environment. The
coupling between a quantum system and its environment leads to
decoherence of the system. Thus, it is natural for us to consider
the process of degradation of entanglement due to the decoherence.
More recently, Yu and Eberly~\cite{Yu} showed that two entangled
qubits become completely disentangled in a finite time under the
influence of pure vacuum noise. Surprisingly, they found that the
behaviors of local decoherence is different from spontaneous
disentanglement. The decoherence effects take an infinite time
evolution under the influence of vacuum while the entanglement
vanishes suddenly in a finite time. Some other researchers also
investigated the process of disentanglement in the open quantum
systems~\cite{Zubairy}-\cite{Roszak}. The problem of decoherence
from spin environments was studied by Cucchietti et al
\cite{Cucchietti}, while they considered the spin environments
consisting of $N$ independent other than correlated spins .

In most of the previous studies, uncorrelated environments are
usually considered, and modelled by a reservoir consists of harmonic
oscillators. Although a collection of harmonic oscillators is a well
approximated modelling to represent the environment weakly coupled
to system, however, in the practical situation, particles in the
environment may have interactions with each other. Consequently, a
problem comes out: How does the entanglement evolves in a correlated
environment? In this paper, we consider this problem and choose a
correlated spin chain, the Ising model in a transverse field, as the
surrounding system. Moreover, this surrounding system displays
quantum phase transition (QPT) at some critical point and thus it
possesses the dynamic hypersensitivity with respect to the
perturbation even induced by a single qubit~\cite{quan}.

As a quantum critical phenomenon, QPT happens at zero temperature,
at which the thermal fluctuations vanish. Thus, QPT is driven only
by quantum fluctuation. Usually, at the critical point there exists
degeneracy between the energy levels of the systems when QPT
happens. Therefore, it can be excepted that, when we study the
dynamic evolution of the system coupled to a environment with QPT,
some special dynamic features will appear at the critical point.
Quan et al~\cite{quan} have studied the decoherence induced by the
correlated environment. It was shown that at the critical point of a
QPT the decoherence is enhanced. Following this work, Cucchietti et
al~\cite{Cucchietti2} discovered that the decoherence induced by the
critical environment possesses some universality with the
Boson-Hubbard model as an illustration.

Now, we consider two spins coupled to the Ising spin chain in a
transverse field, and the purpose is to reveal the effect of the
correlated environment on the dynamic evolution of the two-spin
entanglement. We will study different cases including two qubits and
qutrits. Moreover, we will consider cases that the two spins
initially start from a pure maximally entangled state and a mixed
Werner state~\cite{werner}. The `sudden death' of entanglement is
found to be a quite common phenomenon.

This paper is organized as follows. In Sec.~II, we introduce the
model of two-spin system coupled to Ising spin chain with a
transverse field. By exactly diagonalizing the Hamiltonian, we give
expression of the time
evolution operator. In Sec.~III, the analytical results of the concurrence~%
\cite{Conc} of the two qubits are calculated to show the dynamics of
entanglement. Numerical results are also given to illustrate the details of
the dynamical behaviors of entanglement. In Sec.~IV, two qutrits are coupled
to the Ising spin chain. The analytical and numerical results of the
negativity~\cite{Horodecki,Vidal} are given. At last we give the conclusion
in Sec. V.

\section{Model Hamiltonian and evolution operator}

We choose the engineered environment system to be an Ising spin
chain in a transverse field which displays a QPT. Two spins are
transversely coupled to the chain. The corresponding Hamiltonian
reads
\begin{equation}
H=\sum_{l=-M}^{M}\sigma _{l}^{x}\sigma _{l+1}^{x}+\left[ {\lambda +}\frac{{g}%
}{2}{(s}_{1z}+{s}_{2z}{)}\right] \sum_{l=-M}^{M}\frac{\sigma _{l}^{z}}{2},
\label{hhh}
\end{equation}
where ${\lambda }$ characterizes the strength of the transverse field, ${g}$
denotes the coupling strength between the Ising chain and the two spins, ${s}%
_{1}{\ }$and ${s}_{2}$, $\sigma _{l}^{\alpha }\left( \alpha
=x,y,z\right) $ are the Pauli operators defined on the $l$-th site,
and the total number of spins in the Ising chain is $L=2M+1$. The
Ising model is the simplest model which exhibits a QPT, and can be
exactly calculated.

In order to diagonalize the Hamiltonian, firstly we notice that $\left[ {s}%
_{1z}+{s}_{2z},\sigma _l^\alpha \right] =0,$ thus it is convenient to define
an operator-valued parameter
\begin{equation}
{\hat{\Lambda}}={\lambda +}\frac{{g}}2{(s}_{1z}+{s}_{2z}{),}
\end{equation}
which is a conserved quantity. When we diagonalize the Ising spin
chain, the parameter $\hat{\Lambda }$ can be treated as a $c$-number
with different values corresponding to the eigenvalues of
${s}_{1z}+{s}_{2z}$ in the two-spin subspace.

By combining Jordan-Wigner transformation and Fourier transformation
to the momentum space~\cite{S.Sachdev}, the Hamiltonian can be
written as~\cite{YDWang}
\begin{equation}
H=\sum_{k>0}e^{i\frac{\theta _{k}}{2}\sigma _{kx}}\left( \Omega _{k}\sigma
_{kz}\right) e^{-i\frac{\theta _{k}}{2}\sigma _{kx}}+\left( -\frac{\hat{%
\Lambda }}{2}{+1}\right) \sigma _{0z}  \label{diag_H}
\end{equation}
where we have used the following pseudospin operators $\sigma
_{k\alpha }\left( \alpha =x,y,z\right)$~\cite{YDWang}
\begin{eqnarray}
\sigma _{kx} &=&d_{k}^{\dagger }d_{-k}^{\dagger }+d_{-k}d_{k},\left(
k=1,2,...M\right) \   \notag \\
\sigma _{ky} &=&-id_{k}^{\dagger }d_{-k}^{\dagger }+id_{-k}d_{k},  \notag \\
\sigma _{kz} &=&d_{k}^{\dagger }d_{k}+d_{-k}^{\dagger }d_{-k}-1,  \notag \\
\sigma _{0z} &=&2d_{0}^{\dagger }d_{0}-1,
\end{eqnarray}
and $d_{k}^{\dagger },d_{k} \{k=0,1,2,...\}$\ denote the fermionic
creation and annihilation operators in the momentum space,
respectively. Here,
\begin{eqnarray}
\bigskip \Omega _{k} &=&\sqrt{\left[ -{\hat{\Lambda}+}2\cos \left( 2\pi
k/L\right) \right] ^{2}+4\sin ^{2}\left( 2\pi k/L\right) },  \label{p1} \\
\theta _{k} &=&\arcsin \left[ \frac{-2\sin \left( \frac{2\pi k}{L}\right) }{%
\Omega _{k}}\right] .  \label{p2}
\end{eqnarray}
From Eq.~(\ref{diag_H}) and the units where $\hbar=1$, the time
evolution operator is obtained as:
\begin{equation}
U(t)=e^{-i(-\frac{{\hat{\Lambda}}}{2}{+1)}\sigma _{0z}t}\prod_{k>0}e^{i\frac{%
\theta _{k}}{2}\sigma _{kx}}e^{-it\Omega _{k}\sigma _{kz}}e^{-i\frac{\theta
_{k}}{2}\sigma _{kx}}.  \label{uuu}
\end{equation}
Having explicitly known the evolution operator, we now consider the
entanglement dynamics of the two qubits and two qutrits.

\section{Dynamical disentanglement of two qubits}

\subsection{The case with initial pure entangling state}

We investigate the dynamic evolution of two-qubit entanglement and
assume that the two qubits initially start from a maximally
entangled state.
\begin{equation}
|\Phi \rangle =\frac{1}{\sqrt{2}}\left( \left\vert 00\right\rangle
+\left\vert 11\right\rangle \right) .  \label{mes}
\end{equation}
Here, $\left\vert 0\right\rangle $ and $\left\vert 1\right\rangle $\quad
denote the spin up and down, respectively. The initial state of environment
is assumed to be the vacuum state in the momentum space, namely, $\left\vert
\psi _{E}\right\rangle =|0\rangle _{k=0}\otimes _{k>0}|0\rangle
_{k}|0\rangle _{-k}$, and the vacuum state $|0\rangle _{k}\ $satisfies $%
d_{k}|0\rangle _{k}=0$. We may write a more general initial state of this
composite system as
\begin{equation}
|\Psi (0)\rangle =\left( a\left\vert 00\right\rangle +b\left\vert
11\right\rangle \right) \otimes \left\vert \psi _{E}\right\rangle .
\label{state_tot}
\end{equation}
From the evolution operator (\ref{uuu}), the state vector at time $t$ is
given by
\begin{equation}
|\Psi (t)\rangle =a\left\vert 00\right\rangle \otimes U_{0}\left\vert \psi
_{E}\right\rangle +b\left\vert 11\right\rangle \otimes U_{1}\left\vert \psi
_{E}\right\rangle ,
\end{equation}
where the unitary operator $U_{0}$ and $U_{1}$ can be obtained from the
unitary operator $U(t)$ by replacing operator $\hat{\Lambda }$\quad with number $%
{\lambda +}{{g}}/{2}${\ and }${\lambda -}{{g}}/{2},$ respectively.

Tracing out the environment, in the basis spanned by $\{\left\vert
00\right\rangle ,\left\vert 11\right\rangle ,\left\vert
01\right\rangle ,\left\vert 10\right\rangle \},$ the reduced
density matrix of the two-spin system is obtained as
\begin{equation}
\rho _{1,2}=\left(
\begin{array}{cc}
|a|^2 & ab^{*}F(t) \\
a^{*}bF^{*}(t) & |b|^2%
\end{array}
\right) \oplus Z_{2\times 2},
\end{equation}
where $F(t)=\langle \psi _E|U_1^{\dagger }U_0\left| \psi _E\right\rangle $
is the \emph{decoherence factor}, and $Z_{2\times 2}$ denotes the $2\times 2$
zero matrix. Now, the concurrence~\cite{Conc} of the reduced density matrix
can be readily given by
\begin{equation}
C=2|ab^{\ast }F(t)|=C_{0}|F(t)|,  \label{concurrence}
\end{equation}
where $C_{0}$ is the concurrence of the initial state. We see that
the concurrence is proportional to the norm of the decoherence
factor, and
when the initial state is in a maximally entangled state (\ref{mes}), $%
C=|F(t)|,$ namely, the concurrence is equal to the norm of the
decoherence factor.

Let us consider the decoherence factor
\begin{equation}
F(t)=\langle \psi _{E}|U_{1}^{\dagger }U_{0}\left\vert \psi
_{E}\right\rangle=\prod_{k>0}F_{k},
\end{equation}
where $U_{n}(n=0,1)$ is generated from Hamiltonian $H_{n}$ with $\hat{\Lambda} =$ $%
\Lambda _{n}$(a number). From the unitary operator (\ref{uuu}) and
the initial vacuum state, we obtain
\begin{eqnarray}
|F(t)| &=&\prod_{k>0}\big\{1-\big[\sin( \Omega _{k}^{(0)}t) \cos(
\Omega
_{k}^{(1)}t) \sin \theta _{k}^{(0)}  \notag \\
&&-\cos( \Omega _{k}^{(0)}t) \sin( \Omega _{k}^{(1)}t) \sin \theta _{k}^{(1)}%
\big]^{2}  \notag \\
&&-\sin ^{2}( \Omega _{k}^{(0)}t) \sin ^{2}( \Omega _{k}^{(1)}t) \sin ^{2}(
\theta _{k}^{(0)}-\theta _{k}^{(1)})\big\}^{\frac{1}{2}},  \notag \\
&&  \label{ft}
\end{eqnarray}
where $\Omega _{k}^{(n)}$ and $\theta _{k}^{(n)}$ are obtained by replacing $%
{\hat{\Lambda}}$ with ${\Lambda }_{n}$ in Eqs.~(\ref{p1}) and
(\ref{p2}), respectively. Here, $\Lambda _{0}={\lambda
+}{{g}}/{2}${\ and }$\Lambda _{1}={\lambda -}{{g}}/{2}.$ This is one
of our main results. We see that the zero mode ($k=0$) has no
contribution to the decoherence factor. Clearly, every factor
$F_{k}$ is less than unit. So it can be well expected that in the
large $L$\ limit, $\vert F(t)\vert $ will go to zero under some
reasonable conditions.

By carrying out similar analysis of Ref.~\cite{quan}, we introduce a cutoff number $%
K_{c}$\ and define the partial product for the decoherence factor
\begin{equation}
\left\vert F(t)\right\vert _{c}=\prod_{k>0}^{K_{c}}F_{k}\geq \left\vert
F(t)\right\vert ,
\end{equation}
from which the corresponding\ partial sum
\begin{equation}
S\left( t\right) =\ln \left\vert F(t)\right\vert _{c}\equiv
-\sum_{k>0}^{K_{c}}\left\vert \ln F_{k}\right\vert .
\end{equation}
For the case of small $k$ and large $L$, we have\ $\Omega _{k}^{(n)}\approx
\left\vert 2-\Lambda _{n}\right\vert $, consequently
\begin{equation}
\sin ^{2}\left( \theta _{k}^{(0)}-\theta _{k}^{(1)}\right) \approx \frac{%
16k^{2}\pi ^{2}\left( \Lambda _{0}-\Lambda _{1}\right) ^{2}}{L^{2}\left(
2-\Lambda _{0}\right) ^{2}\left( 2-\Lambda _{1}\right) ^{2}}.
\end{equation}
As a result, if $L$\ is large enough and $\Lambda _{0}-\Lambda _{1}$ is very
small perturbation the\ approximation of $S$ can be obtained as
\begin{eqnarray}
S\left( t\right) &\approx &-2E\left( K_{c}\right) \left( 2-\Lambda
_{0}\right) ^{-2}\left( 2-\Lambda _{1}\right) ^{-2}  \notag \\
&&\times \{\left( \Lambda _{0}-\Lambda _{1}\right) ^{2}\sin ^{2}\left(
\left\vert 2-\Lambda _{0}\right\vert t\right) \sin ^{2}\left( \left\vert
2-\Lambda _{1}\right\vert t\right)  \notag \\
&&+[\sin \left( \left\vert 2-\Lambda _{0}\right\vert t\right) \cos \left(
\left\vert 2-\Lambda _{1}\right\vert t\right) |2-\Lambda _{1}|  \notag \\
&&-\sin \left( \left\vert 2-\Lambda _{1}\right\vert t\right) \cos \left(
\left\vert 2-\Lambda _{0}\right\vert t\right) |2-\Lambda _{0}|]^{2}\},
\notag \\
&&  \label{S}
\end{eqnarray}
where
\begin{equation}
E\left( K_{c}\right) =4\pi ^{2}K_{c}\left( K_{c}+1\right) \left(
2K_{c}+1\right) /\left( 6L^{2}\right) .  \label{E_Kc}
\end{equation}
In the derivation of the above equation, we have used $\ln (1-x)\approx -x$%
\quad for small $x$ and $\sum\limits_{k=1}^{n}k^{2}=n(n+1)(2n+1)/6.$

For our two-qubit case, $\Lambda _{0}={\lambda +}{{g}}/{2}${,
}$\Lambda _{1}={\lambda -}{{g}}/{2}.$ When $\lambda \rightarrow 2$,
and with a proper small $g$\ we have
\begin{equation}
\left\vert F(t)\right\vert _{c}\approx e^{-\gamma t^{4}}  \label{ft_cut}
\end{equation}
with $\gamma =2E\left( K_{c}\right) g^{2}.$ Notice that $\left\vert
F(t)\right\vert _{c}$ is larger than $\left\vert F(t)\right\vert =C.$
Therefore, from the above heuristic analysis we may expect that when the
parameter ${\lambda }$ is adjusted to the vicinity of the critical point $%
\lambda _{c}=2, $ the concurrence (or the decoherence factor) will
exponentially decay with the fourth power of time. Moreover, for short
times, from Eq.~(\ref{ft}), the concurrence becomes
\begin{equation}
C\approx e^{-\Gamma t^{4}}  \label{c_cut}
\end{equation}
with $\Gamma =1/2\sum\limits_{k>0}\sin ^{2}( \theta _{k}^{(0)}-\theta
_{k}^{(1)}) ( \Omega _{k}^{(0)}) ^{2}( \Omega _{k}^{(1)}) ^{2}.$

\begin{figure}[tbp]
\includegraphics[bb=9 474 572 759, width=9 cm, clip]{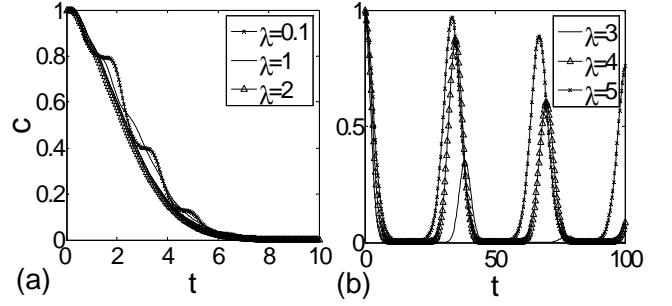}
\caption{\textbf{(a)} Concurrence versus time $t$ with different $\protect%
\lambda$ in the case of weak coupling strength $g=0.1$. The size of
the environment is $L=300$. \textbf{(b)} shows the cases of larger $\protect%
\lambda$. }
\end{figure}

\begin{figure}[tbp]
\includegraphics[bb=66 230 501 595, width=6 cm, clip]{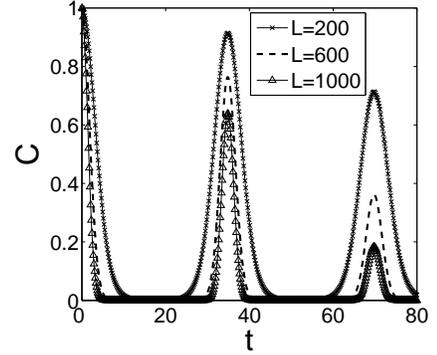}
\caption{Concurrence versus time with different environment size
$L=200, 600$ and $1000$. The transverse field $\protect\lambda=4$,
and the coupling strength $g=0.1$.}
\end{figure}

\begin{figure}[tbp]
\includegraphics[bb=64 247 500 573, width=6 cm, clip]{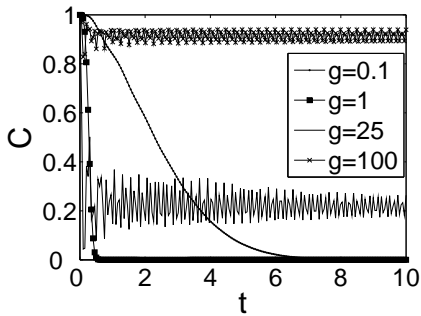}
\caption{Concurrence versus time at the critical point $\protect\lambda =2$
with different coupling strength $g$.}
\end{figure}
Now we resort to numerical analysis of the dynamical sensitivity and
the concurrence decay. In the Fig.~1 \textbf{(a)} and \textbf{(b)},
we plot the concurrence versus time for different $\lambda $. We
find that in the vicinity of the critical point about $\lambda \in
\lbrack 2-0.3,2+0.3]$, concurrence decays monotonously with time.
And extending the time range, however there are not the revivals of
concurrence. Figure 1 \textbf{(a)} shows the cases of $\lambda \leq
2$. We can see that concurrence for the case $\lambda =2$ decays
more rapidly than other cases. It should be noted that, the dynamics
of the two-qubit entanglement in Eq.~(\ref{concurrence}) is
absolutely determined by the decoherence factor in Eq.~(\ref{ft}),
thus from a theoretical point of view, the complete disentanglement
cannot be realized in a finite time. When parameter $\lambda $
becomes larger than $\lambda _{c}$,($g=3,4$ and $5$), the numerical
results of the concurrence are shown in Fig. 1 \textbf{(b)}. The
concurrence oscillates with time, and collapses and revivals are
observed. This is in contrast with the case of small $\lambda$,
where no revivals are found.

The surrounding system displays a QPT near the critical point, and
there exists a competition between different order tendencies
~\cite{S.Sachdev}. From another point of view, near the critical
point quantum chaotic behaviors may emerge~\cite{Emary}. For a
system with quantum chaos, though it is prepared in identical
initial state, two slightly different interactions can lead to two
quite different quantum evolutions. In our system the decoherence
factor can act as a fidelity and quantify the difference between the
two states which are produced through two different evolutions.
Decay of the fidelity can indicate the presence of the quantum
chaos~\cite{Emerson}, and here the monotonous decay of the
decoherence factor (concurrence) at the critical point may be
considered as a signature of quantum chaos.

In Fig.~2, for weak coupling $g=0.1$ and $\lambda=4$, the
oscillation of concurrence is suppressed by enlarging the size of
environment. The larger environment prevents the revival of
entanglement. In the short-time region, we can see the larger size
of environment will accelerate the monotonous decay of concurrence.
From Eq.~(\ref{ft}), each factor $F_k$ is smaller than 1, thus it is
reasonable that large size of environment will be more effective to
suppress the factor $F(t)$, and consequently suppress the
concurrence.

In Fig.~3, we consider the effects of coupling $g$ on the dynamics of
entanglement. At the critical point $\lambda =2$, we adjust $g$ from a small
one $g=0.1$ to a strong one $g=100$. It can be found that when we properly
enlarge the coupling, e.g. $g=1$, the concurrence decays more sharply than
the case $g=0.1$. However, when we continue enlarging the coupling to about $%
g>10 $, e.g. $g=25$, concurrence will oscillate quickly and does not
decay monotonously to zero any more. For the case of very large
coupling $g=100$, concurrence behaves as a weak oscillation near the
initial value of $C=1$. It can be expect that to the strong coupling
limit of $g$, the concurrence will stay at $C=1$ without changing
with time. The above behaviors remind us of the quantum Zeno effects
in process of quantum measurement~\cite{Koshino}. The phenomena
shown in Fig. 3 is similar to the decay probability which can be
suppressed by the increasing coupling between system and measuring
apparatus in quantum Zeno effects.

\subsection{The case of mixed state}

Now, we study the dynamics of disentanglement of mixed entangled
state and assume the two qubits being initially in a Werner state~
\cite{werner}, which is given by
\begin{equation}
\rho _{s}=P|\Phi \rangle \left\langle \Phi \right\vert +\frac{1-P}{4}%
I_{4\times 4},  \label{werner}
\end{equation}
where $|\Phi \rangle $ is the maximally entangled state given by Eq. (\ref%
{mes}), the parameter $P\in [0,1]$, and $I_{4\times 4}$ denotes a $4\times 4$
identity matrix. This state is a mixed state except the extreme case of $P=1$%
. Only when $P>1/3$, the Werner state $\rho _{s}$ is entangled.

We assume the initial state of the whole system $\rho
_{\text{tot}}$ is in\ a direct product form as
\begin{equation}
\rho_\text{tot}=\rho _s\otimes \left| \psi _E\right\rangle \left\langle \psi
_E\right|,
\end{equation}
where $\left| \psi _E\right\rangle $ is the initial state of the
environment. After the time evolution, we can obtain the reduce density
matrix of the two-qubit system in the basis spanned by $\{\left|
00\right\rangle ,\left| 11\right\rangle ,\left| 01\right\rangle ,\left|
10\right\rangle \}$ as follows
\begin{equation}  \label{state}
\rho _{1,2}=\frac 12\left(
\begin{array}{cc}
\frac{1+P}2 & PF(t) \\
PF^{*}(t) & \frac{1+P}2%
\end{array}
\right) \oplus \left( \frac{1-P}4\right) I_{2\times 2},
\end{equation}
where the decoherence factor $F(t)$ is the same as Eq.~(\ref{ft}).

From Eq.(\ref{state}), the concurrence is derived as
\begin{equation}
C=\max \left\{ 0,P\left( \left| F\right| +\frac 12\right) -\frac 12\right\}.
\end{equation}
When $P=1$, it reduces to Eq.~(\ref{concurrence}) for the pure maximally
entangled state. While in the region $1/3<P<1$, the concurrence vanishes
when the decoherence factor
\begin{equation}
\left| F\right| \leq (P^{-1}-1)/2.
\end{equation}
Thus there exists a finite disentanglement time $t_d,$ after which the
entanglement is zero. According to the results of heuristic analysis in Eq.~(%
\ref{ft_cut}), $\left\vert F(t)\right\vert _{c}\approx e^{-\gamma t^{4}}$, in the condition of weak coupling and $\lambda \rightarrow 2$%
, we can approximately give the disentanglement time
\begin{equation}
t_d=\left( \frac 1\gamma \ln \frac{2P}{1-P}\right) ^{\frac 14}.
\end{equation}
Then, the disentanglement time increases as the probability $P$ increases
from 1/3 to 1.

\begin{figure}[tbp]
\includegraphics[bb=75 244 593 572, width=7 cm, clip]{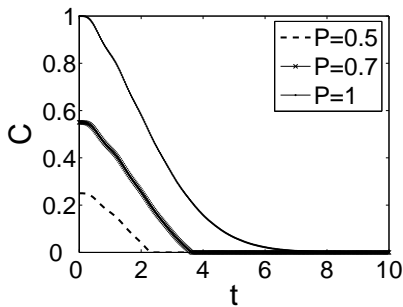}
\caption{Concurrence versus time at the critical point $\protect\lambda =2$
and coupling strength $g=0.1$ for parameters $P=0.5,$ $0.7$ and $1$.}
\end{figure}

In Fig.~4, we also numerically calculate the concurrence versus time for
different probabilities. For the mixed states corresponding to $P=0.5$ and $%
0.7$, disentanglement process takes only a finite time, while for
the pure state case ($P=1$), disentanglement is only completed
asymptotically, and it will take an infinite time. Numerical
results are consistent with the above analytical results that the
disentanglement time increases with the increase of $P$.

\section{Dynamical entanglement evolution of two qutrits}

Now, we consider the case of two qutrits and use the negativity~\cite{Horodecki}
to quantify entanglement. For the systems with spin larger than $%
1/2$, a non-entangled state has necessarily a positive partial transpose
(PPT) according to the Peres-Horodecki criterion~\cite{Horodecki}. In the
case of two spin halves, and the case of (1/2,1) mixed spins, a PPT is also
sufficient. Vidal and Werner~\cite{Vidal} developed the Peres-Horodecki
criterion and presented a measure of entanglement called negativity that can
be computed efficiently, and the negativity does not increase under local
manipulations of the system. The negativity of a state $\rho $ is defined as
\begin{equation}
\mathcal{N(\rho )}=\sum_{i}|\mu _{i}|,
\end{equation}%
where $\mu _{i}$ is the negative eigenvalue of $\rho ^{T_{2}}$, and $T_{2}$
denotes the partial transpose with respect to the second subsystem. If $%
\mathcal{N}>0$, then the two-spin state is entangled. The negativity
has\ been used to characterize the entanglement in large spin system
very well~\cite{Schliemann}-\cite{Zhe}. And by means of negativity,
Derkacz et al. have studied the process of disentanglement in a pair
of three-level atoms interacting with the vacuum~\cite{Derkacz}.

\subsection{The case with initial pure state}

In a similar vein as the study of two-qubit case, we write a
general initial state of the many-body system as
\begin{equation}
|\Psi (0)\rangle =\left( a\left\vert 00\right\rangle +b\left\vert
11\right\rangle +c|22\rangle \right) \otimes \left\vert \psi
_{E}\right\rangle .
\end{equation}%
where $\left\vert 0\right\rangle $, $\left\vert 1\right\rangle ,|2\rangle $
denote the spin-one state with magnetic quantum number 1, 0, -1
respectively. From the evolution operator (\ref{uuu}), the state vector at
time $t$ is given by
\begin{eqnarray}
|\Psi (t)\rangle  &=&a\left\vert 00\right\rangle \otimes U_{0}\left\vert
\psi _{E}\right\rangle +b\left\vert 11\right\rangle \otimes U_{1}\left\vert
\psi _{E}\right\rangle   \notag \\
&&+c|22\rangle \otimes U_{2}\left\vert \psi _{E}\right\rangle ,
\end{eqnarray}%
where the unitary operator $U_{0}$, $U_{1},$and $U_{2}$ are obtained from
the unitary operator $U(t)$ by replacing operator ${\hat{\Lambda}}$ with
number ${\lambda +}g${, }${\lambda }$ and ${\lambda }-g,$ respectively.

In the basis spanned by $\{\left\vert 00\right\rangle $, $\left\vert
11\right\rangle $, $|22\rangle $, $\left\vert 01\right\rangle $, $\left\vert
10\right\rangle $, $\left\vert 02\right\rangle $, $\left\vert
20\right\rangle $, $\left\vert 12\right\rangle $, $\left\vert
21\right\rangle \}$, the reduced density matrix of the two-qutrit system is

\begin{eqnarray}
\rho _{1,2} &=&\left(
\begin{array}{ccc}
|a|^2 & ab^{*}F_1(t) & ac^{*}F_2(t) \\
a^{*}bF_1^{*}(t) & |b|^2 & bc^{*}F_3(t) \\
a^{*}cF_2^{*}(t) & b^{*}cF_3^{*}(t) & |c|^2%
\end{array}
\right) \oplus  \notag \\
&&\oplus Z_{2\times 2}\oplus Z_{2\times 2}\oplus Z_{2\times 2},
\end{eqnarray}
where
\begin{eqnarray}
F_1(t) &=&\langle \psi _E|U_1^{\dagger }U_0\left| \psi _E\right\rangle ,
\notag \\
F_2(t) &=&\langle \psi _E|U_2^{\dagger }U_0\left| \psi _E\right\rangle ,
\notag \\
F_3(t) &=&\langle \psi _E|U_2^{\dagger }U_1\left| \psi _E\right\rangle
\label{ft_1}
\end{eqnarray}
are the decoherence factors.

The partial transpose with respect to the second system gives
\begin{equation}
\rho _{1,2}^{T_{2}}=\text{diag}(|a|^{2},|b|^{2},|c|^{2})\oplus B_{1}\oplus
B_{2}\oplus B_{3},
\end{equation}%
where the three $2\times 2$ matrices
\begin{eqnarray}
B_{_{1}} &=&\left(
\begin{array}{cc}
0 & ab^{\ast }F_{1}(t) \\
a^{\ast }bF_{1}^{\ast }(t) & 0%
\end{array}%
\right) ,  \notag \\
B_{2} &=&\left(
\begin{array}{cc}
0 & ac^{\ast }F_{2}(t) \\
a^{\ast }cF_{2}^{\ast }(t) & 0%
\end{array}%
\right) ,  \notag \\
B_{_{3}} &=&\left(
\begin{array}{cc}
0 & bc^{\ast }F_{3}(t) \\
b^{\ast }cF_{3}^{\ast }(t) & 0%
\end{array}%
\right) .
\end{eqnarray}%
Then, from the above matrix $\rho _{1,2}^{T_{2}}$, one can obtain the
negativity as
\begin{equation}
\mathcal{N}=|ab^{\ast }F_{1}(t)|+|ac^{\ast }F_{2}(t)|+|bc^{\ast }F_{3}(t)|.
\end{equation}%
For the maximally entangled state, $a=b=c=1/\sqrt{3}$, and the negativity
simplifies to
\begin{equation}
\mathcal{N}=\frac{1}{3}\left( |F_{1}(t)|+|F_{2}(t)|+|F_{3}(t)|\right) .
\label{negativity}
\end{equation}%
From the above equation, we can find the negativity is a linear
combination of three decoherence factors. Also with the vacuum state
of environment, the decoherence factors $|F_{\nu }(t)|=\langle \psi
_{E}|U_{j}^{\dagger }U_{i}\left\vert \psi _{E}\right\rangle $ are
given by Eq.(\ref{ft}) by the replacements $\Omega
_{k}^{(0)}\rightarrow \Omega _{k}^{(i)},\Omega _{k}^{(1)}\rightarrow
\Omega _{k}^{(j)},\theta _{k}^{(0)}\rightarrow \theta
_{k}^{(i)},\theta _{k}^{(1)}\rightarrow \theta _{k}^{(j)}.$  Here,
$F_{\nu
}(t)$ denotes the three factors $F_{1}(t),$ $F_{2}(t)$ and $F_{3}(t).$ $%
U_{j}^{\dagger }U_{i}$ correspond to $U_{1}^{\dagger }U_{0},U_{2}^{\dagger
}U_{0}$ and\ $U_{2}^{\dagger }U_{1}$ in\ the three factors Eq.~(\ref{ft_1}).
The parameters $\Omega _{k}^{(n)}$ and $\theta _{k}^{(n)}(n=0,1,2)$ can be
obtained by substituting ${\Lambda }_{0}={\lambda +}g$, ${\Lambda }_{1}={%
\lambda }$ and ${\Lambda }_{2}={\lambda -}g$ into Eq.~(\ref{p1}) and (\ref%
{p2}).

During the similar analysis in the case of two qubits, we can also introduce
the cutoff number $K_{c}$\ and define the partial product for the three\
decoherence factors. Through the small $k$ approximation, we can obtain the
three partial sums corresponding to the three factors. Therefore, under the
condition of weak coupling $g$ and $\lambda \rightarrow 2,$ in a finite time
the three factors $F_{1}(t)$, $F_{2}(t)$ and\ $F_{3}(t)$\ will decay
exponentially with time in a similar form as Eq.~(\ref{ft_cut}).
\begin{figure}[tbp]
\includegraphics[bb=20 327 557 605, width=9 cm, clip]{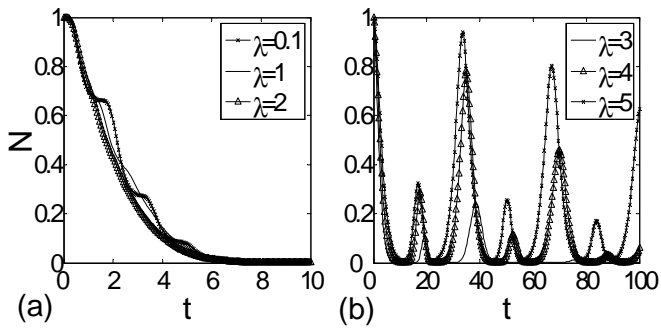}
\caption{\textbf{(a)} Negativity versus time with different cases of $%
\protect\lambda =0.1,1$ and $2$. The coupling $g=0.1$ and the size
of environment  $L=300$. \textbf{(b)} shows the cases of
$\protect\lambda =3,4 $ and $5$. The highest one (solid line with
up triangles) corresponds to the case $\protect\lambda =5$, and
the lowest one (dashed line with points) corresponds to
$\protect\lambda =3$. }
\end{figure}
\begin{figure}[tbp]
\includegraphics[bb=72 250 505 577, width=6 cm, clip]{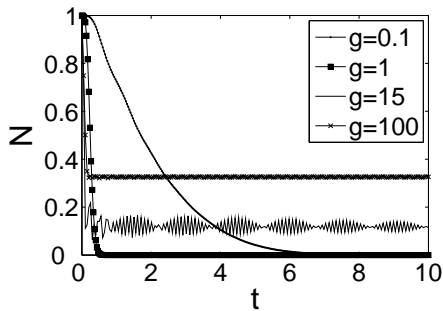}
\caption{ Negativity versus time with different coupling strengths $%
g=0.1,1,15$ and $100$ at the critical point $\protect\lambda _{c}=2$. }
\end{figure}

We numerically calculate the dynamics of negativity. In Fig.~5 \textbf{%
(a)}, it shows the similar phenomena in Fig.~1 \textbf{(a)}. When
the coupling $g$ is weak and $\lambda \rightarrow
2,$ the dynamical behaviors of the three decoherence factors in negativity (%
\ref{negativity}) are nearly identical. Each of the factors decay
with time just as in Eq.~(\ref{ft_cut}), thus
it can be understood that negativity also decays monotonously with time in the vicinity of $%
\lambda =2$. In Fig.~5 \textbf{(b)}, we consider the cases of
larger couplings. Comparing it with Fig.~1 \textbf{(b)}, the
behaviors of negativity have some differences with concurrence.
More revivals are found in the behavior of the negativity, and
they result from the linear superposition of the three decoherece
factors.

In Fig.~6, we numerically study the effects of different couplings
$g$ on the dynamics of negativity. Similar to the dynamic
behaviors of the concurrence. With a properly large coupling such
as $g=1$, the decay of negativity will be much sharper. But very
strong coupling ($g=15$) will make negativity oscillate rapidly.
To the strong coupling limit case of $g=100$,
negativity decays from the initial value $\mathcal{N}=1$ to a steady value $%
1/3$, which is different from the concurrence of the two qubits. Let
us carry out the approximate analysis just like in the case of two
qubits. We can obtain three partial sum $S_{1}$, $S_{2}$ and
$S_{3},$ corresponding to the three decoherence factors in
Eq.~(\ref{ft_1}), which are similar to Eq.~(\ref{S}).
When $g\rightarrow \infty $ and $\lambda \rightarrow 2$, we have $%
S_{2}\rightarrow 0$ and $S_{1}=S_{3}\approx -2E\left( K_{c}\right)
t^{2}$\ where$\ E\left( K_{c}\right) $ is in Eq.~(\ref{E_Kc}), thus
negativity will decay sharply to a steady value of $1/3$. We can see
that different dynamic properties of the factors cause the behaviors
of negativity shown in Fig.~6 is different from concurrence in
Fig.~3.

\subsection{The case of mixed state}
We then consider the mixed state, namely, the two-qutrit Werner
state
\begin{equation}
\rho _{s}=P|\Phi\rangle \langle \Phi\vert +\frac{%
1-P}{9}I_{9\times 9},
\end{equation}
where $|\Phi\rangle $ is the maximally entangled state of two
qutrits and $|\Phi\rangle =\left( \left\vert 00\right\rangle
+\left\vert 11\right\rangle +|22\rangle \right) /\sqrt{3}.$ Assume
that the whole system is initially in $\rho_{\text{ tot}}=\rho
_{s}\otimes \left\vert \psi _{E}\right\rangle \left\langle \psi
_{E}\right\vert $. After time evolution operator in Eq.~(\ref{uuu}),
we can obtain the reduce density matrix of the two qutrits at
arbitrary time $t$. Then, we make the partial transpose with respect
to the second system on the reduce density matrix, and obtain
\begin{eqnarray}
\rho _{1,2}^{T_{2}} &=&\frac{1}{9}\text{diag}(1+2P,1+2P,1+2P)  \notag \\
&&\oplus B_{1}\oplus B_{2}\oplus B_{3},
\end{eqnarray}%
where the three $2\times 2$ matrices
\begin{equation}
B_{_{k}} =\frac{1}{3}\left(
\begin{array}{cc}
\frac{1-P}{3} & PF_{k}(t) \\
PF_{k}^{\ast }(t) & \frac{1-P}{3}%
\end{array}%
\right)~~~k=\{1,2,3\}
\end{equation}

From partially transposed reduced density matrix, the negativity is given by
\begin{equation}
\mathcal{N} =\frac{1}{3}\sum_{k=1}^3\max \left\{ 0,P\left( \left\vert
F_{k}(t)\right\vert +\frac{1}{3}\right) -\frac{1}{3}\right\}.
\end{equation}%
Since $\vert F_{k}(t)\vert\leq 1$, the existence of nonzero negativity needs
the parameter $P$ satisfying the condition $1/4<P\leq 1$. From the above
equation, we can also reads that the disentanglement occurs only when all
the three factors satisfy $\left\vert F_{k}(t)\right\vert \leq (P^{-1}-1)/3$.

Furthermore, we study the case of a $d$-dimension Werner state being
the initial state. Thus we give the initial state of the system as
\begin{equation}
\rho _s=\frac Pd\sum_{i,j=0}^{d-1}\left| ii\right\rangle \left\langle
jj\right| +\frac{1-P}{d^2}I_{d^2\times d^2},
\end{equation}
where the basis vector $\left| ii\right\rangle $ is the eigenvector of $%
s_z=s_{1z}+s_{2z}$ with the eigenvalue $2i+1-d$. Then the initial
state of the whole system is also performed by\ a direct product
form as $\rho_{\text {tot}}=\rho _s\otimes \left| \psi
_E\right\rangle \left\langle \psi _E\right| .$ After the similar
process mentioned in the former parts, we have the matrix $\rho
_{1,2}^{T_2}$ denoting the reduce density matrix after the partial
transpose over the second subsystem at time $t$, which is shown as:
\begin{eqnarray}
\rho _{1,2}^{T_2} &=&\frac Pd\sum_{i,j=0}^{d-1}\left| ij\right\rangle
\left\langle ji\right| F_{i,j}(t)+\frac{1-P}{d^2}I_{d^2\times d^2}  \notag \\
&=&\frac 1{d^2}{\text{diag}}\left[ 1+(d-1)P,...,1+(d-1)P\right]
_{d\times d}  \notag
\\
&&\oplus _{i<j}\frac 1d\left(
\begin{array}{cc}
\frac{1-P}d & PF_{i,j}(t) \\
PF_{i,j}^{*}(t) & \frac{1-P}d%
\end{array}
\right) ,
\end{eqnarray}
where the decoherence factors $F_{i,j}(t)=\langle \psi _E|U_j^{\dagger
}U_i\left| \psi _E\right\rangle ,$ and the corresponding time evolution
operator $U_i$ can be obtained from Eq.~(\ref{uuu}) by replacing operator ${%
\hat{\Lambda}}$ with value ${\lambda +}{{g}}/2(2i+1-d),$
respectively. It is apparent that we should only focus on the
$2\times 2$ matrices and obtain the negativity as
\begin{equation}
\mathcal{N}=\frac 1d\sum_{i<j}\max \left\{ 0,P\left( \left|
F_{i,j}(t)\right| +\frac 1d\right) -\frac 1d\right\} ,
\end{equation}
from which we can see that negativity will be complete vanishes when
all the norms satisfy $\left| F_{i,j}(t)\right| \leq (P^{-1}-1)/d$
simultaneously.

\section{conclusion}

In summary, we have studied the dynamics of entanglement in a pure
dephasing system. By making use of the concept of concurrence, we
studied two qubits coupled to an Ising spin chain in a transverse
field. When the two qubits initially started from a pure entangled
state, we obtained the analytical results of concurrence which is
just a simple product of the initial concurrence $C(0)$ and the
decoherence factor $F(t)$. Thus the dynamic properties of
concurrence is absolutely determined by the decoherence factor.
Specially, in the case of weak coupling, the concurrence decays
exponentially with time when $\lambda \rightarrow \lambda _{c}$.
Moreover, we found the decay of decoherence factor is of the form
$\exp(-\Gamma t^4)$, which is not a Gaussian form like in Ref.~\cite%
{quan} and~\cite{Cucchietti2}. Certainly this is due to the initial
state of the environment we have chosen.

Furthermore, when the two qubits are initially in the Werner state,
we have found that the complete disentanglement takes place in a
finite time just as the `sudden death' of entanglement discovered in
Ref.~\cite{Yu}. In \cite{Yu}, due to the process of spontaneous
emission, the sudden death of entanglement can occur in an arbitrary
entangled state (pure or mixed). However, in our system with
dephasing effects, when the two entangled qubits are in a pure
state, there does not exist such a phenomena.

We also considered two qutrits coupled to the Ising spin chain. When
the qutrits initially start from a pure state, we have obtained the
expression of negativity which is a linear combination of three
decoherence factors. With weak coupling, negativity also decays
monotonously in the condition $\lambda \rightarrow 2$. When the
qutrits are initially in a Werner state, the complete
disentanglement could occur in a finite time, and then the
properties of negativity are the three decoherence factors. Indeed,
the correlated environment, especially when QPT happens, greatly
affects the decoherence and the disentanglement process. The
entanglement decay in other environment which displays a
QPT~\cite{Lambert}, or quantum chaos~\cite{Fujisaki} deserves
further investigations.

\acknowledgements This work is supported by NSFC with grant
Nos.10405019 and 90503003; NFRPC with grant No. 2006CB921206;
Specialized Research Fund for the Doctoral Program of Higher
Education (SRFDP) with grant No.20050335087.

\end{document}